\documentclass{jpsj2}
\title{Analysis of Incident-Photon-Energy and Polarization Dependent \\
 Resonant Inelastic X-Ray Scattering from La$_{2}$CuO$_{4}$ }

\author{Manabu Takahashi $^{1}$ \thanks{mtakahas@phys.sci.gunma-u.ac.jp}
      , Junichi Igarashi $^{2}$
       and Takuji Nomura $^{3}$
       }

\inst{$^{1}$Faculty of Engineering, Gunma University, Kiryu, Gunma 376-8515 \\
$^{2}$Faculty of Science, Ibaraki University, Mito, Ibaraki 310-8512 \\
$^{3}$Synchrotron Radiation Research Center, Japan Atomic energy Agency, Hyogo 679-5148}

\abst{We present a detailed analysis of the incident-photon-energy and
 polarization dependences of the resonant inelastic x-ray scattering (RIXS)
 spectra at the Cu $K$ edge in La$_{2}$CuO$_{4}$. Our analysis is based on
 the formula developed by Nomura and Igarashi, which describes the spectra
 by a product of an incident-photon-dependent factor and a density-density
 correlation function for $3d$ states. We calculate the former factor using
 the $4p$ density of states from an ab initio band structure calculation and
 the latter using a multiorbital tight-binding model within the Hartree-Fock
 approximation and the random phase approximation.  We obtain spectra
 with rich structures in the energy-loss range $2$-$5$ eV, which vary with varying momentum and
 incident-photon energy, in semi-quantitative agreement with recent
 experiments. We clarify the origin of such changes as a combined effect of
 the incident-photon-dependent factor and the density-density correlation
 function. 
}

\kword{$\mathrm{La_2CuO_4}$, RIXS, XES, tight-binding model, theory}

\begin{document}
\maketitle

\section{Introduction\label{sec:Introduction}}

Taking advantage of strong synchrotron light sources, the resonant
inelastic x-ray scattering (RIXS) has recently become a powerful tool
to probe charge excitations in solids.
$K$-edge resonances are widely used in transition-metal compounds,
because it could detect momentum dependence of charge excitations.
\cite{Hill1998,Abbamonte1999,Hasan2000,Hasan2002,Kim2002,Inami2003,Kim2004PRB,Kim2004PRL,Suga2005,Ishii2005-1,Ishii2005-2,Lu2005,Lu2006,Collart2006,Ellis2007,Kim2007}
It is described as a second-order optical process that a $1s$-core
electron excites to an empty $4p$ state with absorbing a photon,
then charge excitations are created in the $3d$ states to screen
the core-hole potential, and finally the photoexcited $4p$ electron
recombines with the $1s$-core hole with emitting a photon. In the
end, the charge excitations are left with carrying the energy and
the momentum transferred from photon.

Intensive studies have been carried out on Cu oxide perovskite compounds
for better understanding of the unconventional high-$T_{\mathrm{C}}$
superconductivity. \cite{Hasan2000,Hasan2002,Kim2002,Kim2004PRB,Kim2004PRL,Ishii2005-1,Ishii2005-2,Suga2005,Lu2005,Lu2006,Collart2006,Ellis2007,Kim2007}
In the undoped material La$_{2}$CuO$_{4}$, the spectra are found
to be composed of several peaks which changes with changing momentum
transfer. \cite{Kim2002,Lu2006} Such spectra have been theoretically
studied by several groups. \cite{Tsutsui1999,Tsutsui2000,Nomura2004,Nomura2005,Okada2006,Igarashi2006}
Nomura and Igarashi (NI)\cite{Nomura2004,Nomura2005} have proposed
a general formalism of the RIXS spectra by extending the resonant
Raman theory developed by Nozi\`{e}res and Abrahams\cite{Nozieres1974}
on the basis of the many-body formalism of Keldysh. The NI formula
makes it possible to calculate the RIXS spectra on complicated models
including many orbitals, and provides clear physical interpretations
to the RIXS spectra. It is composed of two factors, one describing
an incident-photon dependence and the other the density-density correlation
function for the $3d$ states. Similar formulas which are proportional
to the density-density correlation function have been derived by using
different methods.\cite{Brink2006}

This formula has been applied to a two-dimensional cuprate La$_{2}$CuO$_{4}$\cite{Nomura2005}
and quasi-one-dimensional cuprates SrCuO$_{3}$\cite{Nomura2004}
and CuGeO$_{3}$.\cite{Suga2005} In these studies, the electronic
structures have been calculated on the $d$-$p$ model within the
Hartree-Fock approximation (HFA) in the antiferromagnetic (AFM) phase.
It is known that the HFA works well to describe electronic structures
in the AFM insulators. Two-particle correlations have been taken into
account within the random phase approximation (RPA). The calculated
spectra as a function of energy loss have reproduced well the RIXS
spectra varying with varying momentum. The RIXS spectra have been
interpreted as a band-to-band transition. The RPA correlation had
to be included for obtaining better agreement with the experiment.

Subsequently, the present authors have analyzed multiple-scattering
contributions due to the core-hole potential, because the core-hole
potential is not definitely weak.\cite{Igarashi2006} Having evaluated
the multiple-scattering contributions by means of the time-representation
method by Nozi\`{e}res and De Dominicis, \cite{Nozieres1969_1097}
we have found that main contributions could be absorbed into the shift
of the core-level energy with minor effects on the RIXS spectral shape.
This result partly justifies the use of the Born approximation. Quite
recently, we have demonstrated the usefulness of the NI formula by
analyzing the RIXS spectra in NiO.\cite{Takahashi2007NiO} Using the
HFA and the RPA on the multiorbital tight-binding model, we have obtained
the RIXS spectra, which vary with varying momentum, in quantitative
agreement with the experiment.\cite{H.Ishii2006}

Recently, Lu et al.\cite{Lu2006} have presented comprehensive RIXS
data with systematically changing the incident-photon energy in La$_{2}$CuO$_{4}$.
Kim et al.\cite{Kim2007} have recounted the momentum transfer dependence
and incident photon polarization dependence of the RIXS line shape in a number of cuprates.
Ellis et al.\cite{Ellis2007} have reported the high resolution RIXS spectra in La$_{2}$CuO$_{4}$.
Their data show fine structures as a function of energy loss, in addition
to the peaks found in previous data, which vary with varying the incident-photon
energy.
Although our previous analysis using the $d$-$p$ model has
been successful to clarify the origin of the RIXS spectra,\cite{Nomura2005}
the analyses have been limited to the situation of the incident-photon
energy corresponding to the excitation of the $1s$ electron to the
peak of the $4p$ density of states (DOS). Also, it seems hard to explain the newly observed
fine structures. Encouraged by the success in the application to NiO,\cite{Takahashi2007NiO}
we extend the $d$-$p$ model to a multiorbital tight-binding model,
which includes all O $2p$ and Cu $3d$ orbitals as well as the full
intra-atomic Coulomb interaction between $3d$ orbitals, and analyze
the incident-photon-energy and polarization dependences of the RIXS
spectra in comparison with the experiment.\cite{Lu2006,Ellis2007}

In the present study, we obtain the AFM insulating
solution having an energy gap $\sim1.7\,\mathrm{eV}$ within the HFA.
Note that the band structure calculation with the local density approximation
(LDA) fails to reproduce the insulating state. These energy bands
by the HFA are used to calculate the density-density correlation function.
We treat the two-particle correlations within the RPA. The incident-photon-dependent
factor is calculated by using the $4p$ DOS obtained
from the ab-initio band structure calculation. We obtain richer structures
in RIXS spectra in a range of energy loss $2\sim5\,\mathrm{eV}$ than those obtained by the $d$-$p$ model analysis\cite{Nomura2005}, which
vary with varying momentum transfer and incident-photon energy. The
obtained spectra are in semi-quantitative agreement with the experiments.\cite{Lu2006,Kim2007,Ellis2007}
We clarify the origin of such spectral change as a combined effect
of the incident-photon-dependent factor and the density-density correlation
function.

The present paper is organized as follows. In Sec. \ref{sec:multi-orbital-tight-binding},
we introduce the multiorbital tight-binding model. In Sec. \ref{sec:Electronic-Structure-within}
we discuss the electronic structure within the HFA in the AFM phase
of La$_{2}$CuO$_{4}$. In Sec. \ref{sec:Formula-for-RIXS}, we briefly
summarize the NI formula for the RIXS spectra. In Sec. \ref{sec:Calculated-Results},
we present the calculated RIXS spectra in comparison with the experiment.
The last section is devoted to the concluding remarks.

\section{Electronic Structure of $\mathrm{La_{2}CuO_{4}}$}

\subsection{Multiorbital tight binding model \label{sec:multi-orbital-tight-binding}}

\begin{figure}[t]
\begin{center}
 \includegraphics[clip,scale=0.6]{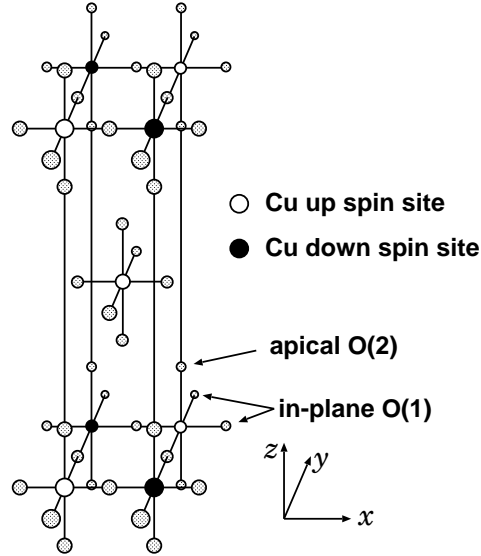}
\end{center}

\caption{Sketch of the assumed crystal and magnetic structure of La$_{2}$CuO$_{4}$
with omitting La sites. The tiny tilting, rotation and the Jahn-Teller
distortion of CuO$_{6}$ octahedron are neglected. \label{fig:crys}}
\end{figure}

For explaining the fine structures in the observed RIXS spectra, we
need to extend a simple $d$-$p$ model to a multiorbital tight-binding
model, in which most of parameters are determined from an ab-initio
band structure calculation.

As shown in Fig.~\ref{fig:crys}, we approximate the crystal structure
of La$_{2}$CuO$_{4}$ by a simple tetragonal one without the Jahn-Teller
distortion and the tilt of the CuO$_{6}$ octahedron. Extending the
$d$-$p$ model, we introduce a tight-binding model involving all
Cu $3d$ orbitals and O $2p$ orbitals at the in-plane and apical
sites. We exclude orbitals belonging to the La atoms, since those
orbitals play minor roles in the electronic states near the insulating
gap. \cite{DeWeert1989} Thereby the Hamiltonian may be written as
\begin{eqnarray}
H & = & H_{0}+H_{\mathrm{I}},\\
H_{0} & = & \sum_{im\sigma}E_{im\sigma}^{d}n_{im\sigma}^{d}+\sum_{jm\sigma}E_{jm\sigma}^{p}n_{jm\sigma}^{p}+\sum_{\left\langle i,j\right\rangle }\sum_{\sigma mm^{\prime}}\left(t_{im,jm^{\prime}}^{dp}d_{im\sigma}^{\dagger}p_{jm^{\prime}\sigma}+\mathrm{H.c.}\right)\nonumber \\
 & + & \sum_{\left\langle j,j'\right\rangle }\sum_{\sigma mm^{\prime}}\left(t_{jm,j'm'}^{pp}p_{jm\sigma}^{\dagger}p_{j'm'\sigma}+\mathrm{H.c.}\right)+\sum_{\left\langle i,i'\right\rangle }\sum_{\sigma mm'}\left(t_{im,i'm'}^{dd}d_{im\sigma}^{\dagger}d_{i'm'\sigma}+\mathrm{H.c.}\right),\\
H_{I} & = & \frac{1}{2}\sum_{i}\sum_{m_{1}\sigma_{1}}\sum_{m_{2}\sigma_{2}}\sum_{m_{3}\sigma_{3}}\sum_{m_{4}\sigma_{4}}g\left(m_{1}\sigma_{1}m_{2}\sigma_{2};m_{3}\sigma_{3}m_{4}\sigma_{4}\right)d_{im_{1}\sigma_{1}}^{\dagger}d_{im_{2}\sigma_{2}}^{\dagger}d_{im_{4}\sigma_{4}}d_{im_{3}\sigma_{3}}.\end{eqnarray}
 The part $H_{0}$ represents the kinetic energy, where $d_{im\sigma}$
and $p_{jm\sigma}$ denote the annihilation operators of an electron
with spin $\sigma$ in the $3d$ orbital $m$ at Cu site $i$ and
that of an electron with spin $\sigma$ in the $2p$ orbital $m$
at the O site $j$, respectively. Number operators $n_{im\sigma}^{d}$
and $n_{jm\sigma}^{p}$ are defined as $n_{im\sigma}^{d}=d_{im\sigma}^{\dagger}d_{im\sigma}$,
$n_{jm\sigma}^{p}=p_{jm\sigma}^{\dagger}p_{jm\sigma}$. The transfer
integrals, $t_{im,jm^{\prime}}^{dp}$, $t_{jm,j'm'}^{pp}$, and $t_{im,i'm'}^{dd}$
are evaluated from the Slater-Koster (SK) two-center integrals, $(pd\sigma)$,
$(pd\pi)$, $(pp\sigma)$, $(pp\pi)$, $(dd\sigma)$, $(dd\pi)$,
$(dd\delta)$.\cite{Slater1954} We use the SK parameters determined
by the ab-initio band structure calculation.\cite{DeWeert1989} The
part $H_{\mathrm{I}}$ represents the intra-atomic Coulomb interaction
on Cu sites. The interaction matrix element $g\left(\nu_{1}\nu_{2};\nu_{3}\nu_{4}\right)$
($\nu$ stands for spin-orbit$\left(m\sigma\right)$) is written in
terms of the Slater integrals $F^{0}$, $F^{2}$, and $F^{4}$. Among
them, $F^{2}$ and $F^{4}$, which are known to be slightly screened
by solid-state effects, are taken from the cluster model exact diagonalization
analysis of the X-ray photoemission spectroscopy.\cite{Eskes1991}
On the other hand, $F^{0}$ is known to be considerably screened,
so that we regard the value as an adjustable parameter. The Coulomb
interaction on O sites is absorbed into a renormalization of the O
$2p$ level parameters $E_{jm\sigma}^{p}$. The $d$-level position
relative to the $p$-levels is given by the charge-transfer energy
$\Delta$ defined as $\Delta=E_{e_{g}}-E_{p}+9U$ in the $d^{9}$
configuration.\cite{Mizokawa1996} Here $U$ is the multiplet-averaged
$d$-$d$ Coulomb interaction given by $U=F^{0}-\left(2/63\right)F^{2}-\left(2/63\right)F^{4}$.
We treat the charge transfer energy $\Delta$ as an adjustable parameter
in our calculation. We fix the charge transfer energy at $\Delta=2.76\,\mathrm{eV}$
in order that the edge position of the RIXS intensity is consistent
with the experiment. Note that the energy gap is determined by the
charge transfer energy, since the electron system in La$_{2}$CuO$_{4}$
belongs to the charge transfer type insulator. We have checked that
the RIXS spectra do not crucially depend on the values of $F^{0}$.
The parameters used in the calculation are listed in Table \ref{table.1}.

\begin{table}[t]

\caption{\label{table.1} Parameter values for the tight-binding model of
La$_{2}$CuO$_{4}$ in units of eV. SK parameters are taken from ref.
\citen{DeWeert1989}. $E_{e_{g}}-E_{t_{2g}}$ is the crystal
field splitting between the Cu $3d$ $e_{g}$ and $t_{2g}$ states.
$E_{\mathrm{O\left(1\right)}}^{p}$ and $E_{\mathrm{O\left(2\right)}}^{p}$
are the O $2p$ levels for the in-plane and apical oxygen, respectively.
Slater Integral $F^{2}$ and $F^{4}$ are taken from ref. \citen{Eskes1991}. }

\begin{tabular}{crrrrrrrrrrr}
\hline 
&
&
&
&
&
&
&
&
&
&
&
\tabularnewline
Cu&
$F^{0}$~&
~~&
$7.3$, $11.3$&
~~~&
$F^{2}$~&
~~&
$11.41$~~&
~~~&
$F^{4}$~&
~~&
$7.308$~~\tabularnewline
&
$E_{e_{g}}-E_{t_{2g}}$&
&
$1.18$~~&
&
$\Delta$~&
&
$2.71$~~&
&
&
&
\tabularnewline
&
$dd\sigma$~&
&
$0.065$~~&
&
$dd\pi$~&
&
$-0.067$~~&
&
$dd\delta$~&
&
$-0.079$~~\tabularnewline
Cu-O(1)&
$pd\sigma$~&
&
$1.253$~~&
&
$pd\pi$~&
&
$0.853$~~&
&
&
&
\tabularnewline
Cu-O(2)&
$pd\sigma$~&
&
$0.568$~~&
&
$pd\pi$~&
&
$0.377$~~&
&
&
&
\tabularnewline
O(1)&
$pp\sigma$~&
&
$0.586$~~&
&
$pp\pi$~&
&
$-0.384$~~&
&
$E_{\mathrm{O\left(1\right)}}^{p}$&
&
$0$~~\tabularnewline
O(1)-O(2)&
$pp\sigma$~&
&
$-0.207$~~&
&
$pp\pi$~&
&
$-0.196$~~&
&
&
&
\tabularnewline
O(2)&
$pp\sigma$~&
&
$0.171$~~&
&
$pp\pi$~&
&
$-0.025$~~&
&
$E_{\mathrm{O\left(2\right)}}^{p}$&
&
$0.50$~~\tabularnewline
\hline
\end{tabular}
\end{table}

\subsection{Hartree-Fock approximation\label{sec:Electronic-Structure-within}}

Assuming the AFM ordering, we choose the unit cell that contains two
Cu sites, one of which is a up spin site and another is
a down spin site, and eight O sites, four of which are
the in-plane O(1) sites and the others apical O(2) sites. Then we
solve the Schrodinger equation $H\Psi=E\Psi$ by disregarding the
fluctuation terms in $H_{I}$, {i.e.}, \begin{equation}
H_{I}^{HF}=\frac{1}{2}\sum_{i}\sum_{\nu_{1}\nu_{2}\nu_{3}\nu_{4}}\Gamma^{(0)}\left(\nu_{1}\nu_{2};\nu_{3}\nu_{4}\right)\langle d_{i\nu_{2}}^{\dagger}d_{i\nu_{3}}\rangle d_{i\nu_{1}}^{\dagger}d_{i\nu_{4}},\end{equation}
 where $\Gamma^{(0)}$ is the antisymmetric vertex function defined
by \begin{equation}
\Gamma^{(0)}\left(\nu_{1}\nu_{2};\nu_{3}\nu_{4}\right)=g\left(\nu_{1}\nu_{2};\nu_{3}\nu_{4}\right)-g\left(\nu_{1}\nu_{2};\nu_{4}\nu_{3}\right),\end{equation}
 with the bracket $\left\langle \cdots\right\rangle $ indicating
the ground state average.

\begin{figure}[t]
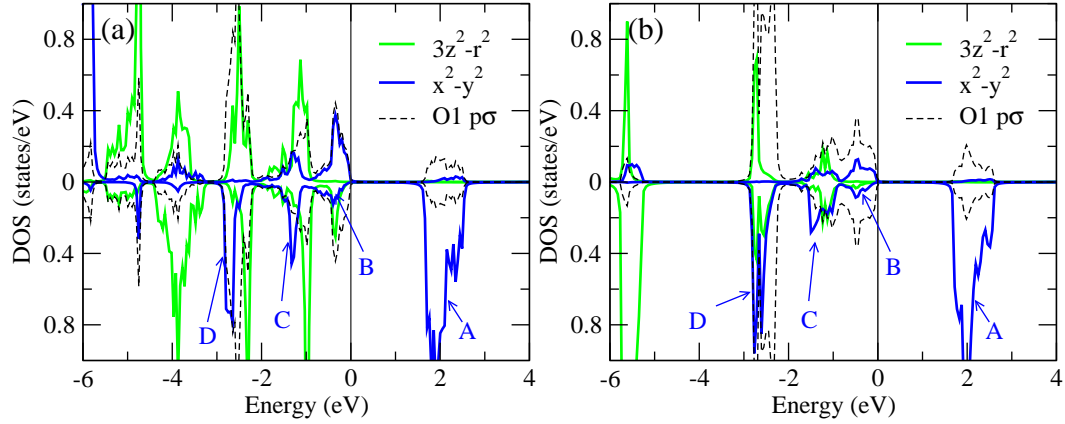

\begin{center}
 \includegraphics[clip,scale=0.55]{62987Fig2a}\includegraphics[clip,scale=0.55]{62987Fig2b}
\end{center}

\caption{\label{fig:dos}(Color online) Density of states projected on the
Cu $3d$ $x^{2}-y^{2}$, $3z^{2}-r^{2}$, and O(1) $2p$ $\sigma$
states. The origin of energy is at the top of valence band. (a) and (b) are for $F^{0}=7.3$ and $11.3$ eV, respectively.}
\end{figure}

\begin{figure}[t]
\begin{center}
\includegraphics[clip,scale=0.6]{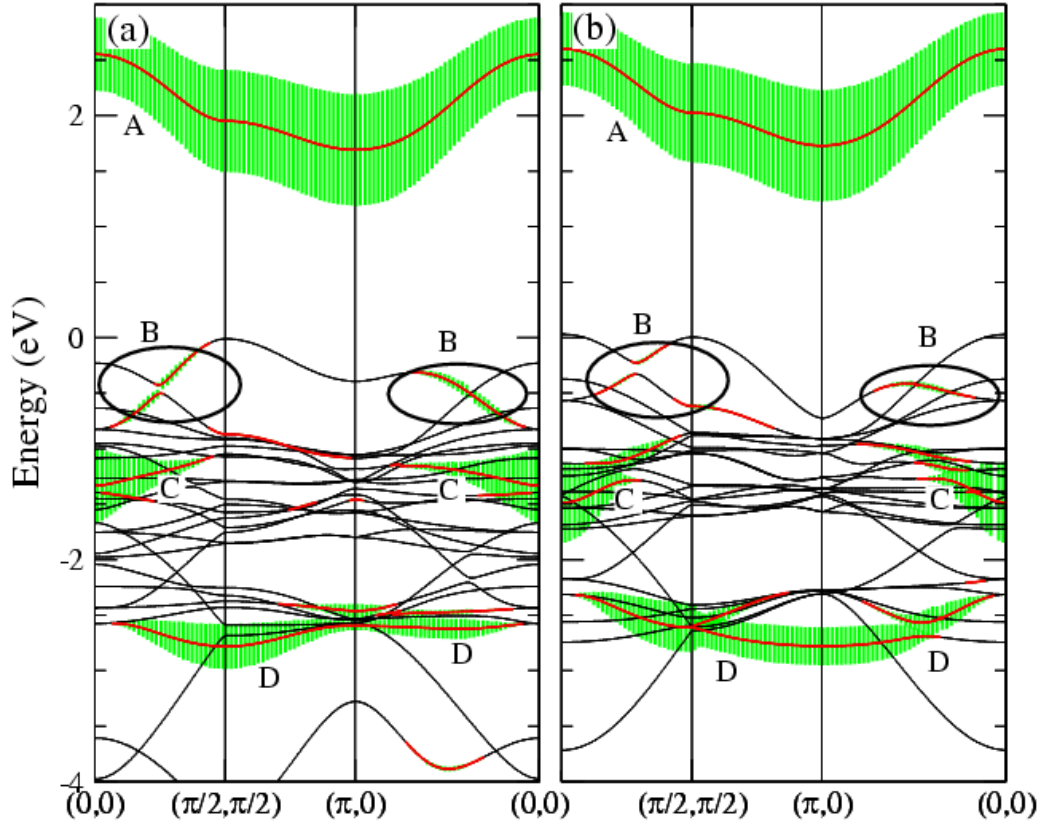}
\end{center}

\caption{\label{fig:disp}(Color online) Dispersion curves along the several
symmetric lines. The weight of minority spin $x^{2}-y^{2}$ states
is shown by the length of the vertical bar. The origin of energy is at the top of valence band. (a) and (b) are for $F^{0}=7.3$
and $11.3$ eV, respectively.}
\end{figure}

For parameter values of $F^{0}$, we take up two typical values $F^{0}=7.3$
and $11.3\,\mathrm{eV}$, which correspond to the values of $U$ used
in refs. \citen{Takahashi1999} and \citen{Nomura2005},
respectively. For both cases, we obtain stable AFM solutions which
have the energy gap $E_{\mathrm{gap}}\sim1.7\,\mathrm{eV}$ and the
spin moment at Cu site $s_{\mathrm{Cu}}\sim0.3\,\hbar$. The fact
that the energy gap depends little on the $F^{0}$ values indicates
that the system belongs to the charge transfer insulating phase. The
density of states and the dispersion curves along some symmetric lines
are shown in Figs. \ref{fig:dos} and \ref{fig:disp}, respectively.
The weight of the minority spin $x^{2}-y^{2}$ states is indicated
by the length of vertical bars in fig. \ref{fig:disp}. Labels assigned
to the DOS peaks and to the dispersion curves will be related to the
peaks in the RIXS intensities.

The states in the conduction band (assigned A) consist mainly of the
minority spin $x^{2}-y^{2}$ state. Thereby the RIXS intensity is
strongly correlated to the minority spin $x^{2}-y^{2}$ state in the
valence band, as discussed in the next section. The states around
the top of valence band (assigned B) are mainly composed of the in-plane
O $p\sigma$ states and the majority spin $x^{2}-y^{2}$ states, but
they also contain small weights of the minority spin $x^{2}-y^{2}$
states, as seen from the figure. Their weights strongly depend on
the momentum $\mathbf{k}$. Particularly, in the valence bands located
just below the gap, the weight is negligible along the $(\pi/2,\pi/2)$-$(\pi,0)$
line, while it becomes discernible along the $(0,0)$-$(\pi/2,\pi/2)$
and $(\pi,0)$-$(0,0)$ lines as denoted by B. The states labeled
C and D noticeably contain the minority spin $x^{2}-y^{2}$ states.
We note that the states relating to the DOS peaks and the dispersion
curves denoted by A, B, C, and D do not strongly depend on the $F^{0}$
values. The energy position and the weight of the $x^{2}-y^{2}$ states
are mainly controlled by the charge transfer energy $\Delta$ and
the SK parameters. We also note that while the dispersion curves for
the states denoted A and B are quite consistent with those calculated
by the $d$-$p$ model, it is difficult to find the correspondence
for the other dispersion curves between the multiorbital model and
the $d$-$p$ model.

\section{Formula for RIXS\label{sec:Formula-for-RIXS}}

We briefly summarize the NI formula for the RIXS, following ref. \citen{Takahashi2007NiO}.
In the RIXS process, the incident photon is absorbed by exciting a
Cu $1s$ core electron to the unoccupied Cu $4p$ state, and a photon
is emitted by recombining the $4p$ electron and the core hole. This
process may be described by \begin{equation}
H_{x}=w\sum_{{\bf q}\alpha}\frac{1}{\sqrt{2\omega_{{\bf q}}}}\sum_{i\eta\sigma}e_{\eta}^{(\alpha)}p_{i\eta\sigma}^{\prime\dagger}s_{i\sigma}c_{{\bf q}\alpha}{\rm e}^{i{\bf q}\cdot{\bf r}_{i}}+{\rm H.c.},\end{equation}
 where $w$ represents the dipole transition matrix element between
the $1s$ and the $4p$ states. We assume that $w$ is constant, since
it changes little in the energy range of $20\,\mathrm{eV}$ above
the absorption edge. The $e_{\eta}^{(\alpha)}$ represents the $\eta$-th
component ($\eta=x,y,z$) of two kinds of polarization vectors ($\alpha=1,2$)
of photon. Annihilation operators $p_{i\eta\sigma}^{\prime}$ and
$s_{i\sigma}$ are for states $4p_{\eta}$ and state $1s$ at Cu site
$i$, respectively. The annihilation operator $c_{{\bf q}\alpha}$
is for photon with momentum ${\bf q}$ and polarization $\alpha$.
In the intermediate state of the RIXS process, the core-hole potential
is acting on the $3d$ states, creating an electron-hole pair. The
interaction is described by \begin{equation}
H_{1s-3d}=V\sum_{im\sigma\sigma'}d_{im\sigma}^{\dagger}d_{im\sigma}s_{i\sigma'}^{\dagger}s_{i\sigma'},\label{eq.h_1s3d}\end{equation}
 where $i$ runs over Cu sites. In the end of the process, an electron-hole
pair is left with momentum and energy $q=\left(\mathbf{q},\omega\right)=\left(\mathbf{q}_{i}-\mathbf{q}_{f},\omega_{i}-\omega_{f}\right)$,
where $q_{i}=\left(\mathbf{q}_{i},\omega_{i}\right)$ and $q_{f}=\left(\mathbf{q}_{f},\omega_{f}\right)$
are energy-momentum of incident and scattered photons.

\begin{figure}[t]
\begin{center}
\includegraphics[clip,scale=0.6]{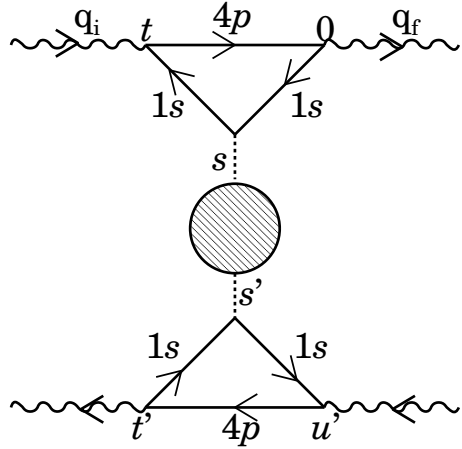}
\end{center}

\caption{\label{fig:diagram1}Diagram for the RIXS intensity within the Born
approximation for the $1s$ core-hole potential. The wavy lines represent
photon Green's functions. The solid lines with the labels $4p$ and
$1s$ represent the bare Green's functions for the $4p$ electron
and the $1s$ core electron, respectively. The dotted lines represent
the core-hole potential $V$. The shaded part represents the density-density
correlation function of the Keldysh-type.}
\end{figure}

The RIXS intensity is derived on the basis of the Keldysh-Green function
scheme. The diagrammatical representation of the process is shown
in Fig.\ref{fig:diagram1} (see ref.\citen{Takahashi2007NiO}
for details). Within the Born approximation to the core-hole potential,
we obtain \begin{equation}
W(q_{i},\alpha_{i};q_{f},\alpha_{f})=\frac{N\left|w\right|^{4}}{4\omega_{i}\omega_{f}}Y_{d}^{+-}(q)\left|\sum_{\eta}e_{i\eta}L_{B}^{\eta}\left(\omega_{i};\omega\right)e_{f\eta}\right|^{2}.\label{eq.general}\end{equation}
 Here $N$ is the number of unit cell, $e_{i\eta}$ ($e_{f\eta}$)
is the $\eta$ component of the polarization vector $\mathbf{e}_{i}$
($\mathbf{e}_{f}$) with $\eta=x,y,z$.

The factor $|\sum_{\eta}e_{i\eta}L_{\mathrm{B}}^{\eta}\left(\omega_{i};\omega\right)e_{f\eta}|^{2}$
describes the incident-photon dependence, which is given by \begin{eqnarray}
L_{B}^{\eta}\left(\omega_{i};\omega\right) & = & \frac{V}{N}\int_{\epsilon_{0}}^{\infty}\frac{\rho_{4p}^{\eta\eta}\left(\epsilon\right){\rm d}\epsilon}{\left(\omega_{i}+\epsilon_{1s}+i\Gamma_{1s}-\epsilon\right)\left(\omega_{i}-\omega+\epsilon_{1s}+i\Gamma_{1s}-\epsilon\right)}.\label{eq.born}\end{eqnarray}
 The $\Gamma_{1s}$ represents the life-time broadening width of the
core-hole state, and $\epsilon_{0}$ indicates the energy at the bottom
of the $4p$ band. The $\rho_{4p}^{\eta\eta'}$ is the DOS matrix
in the $p$ symmetric states, which may be given by \begin{equation}
\rho_{4p}^{\eta\eta'}\left(\epsilon\right)=\sum_{\sigma}\sum_{j\mathbf{k}}\phi_{\eta\sigma j}^{*}\left(\mathbf{k}\right)\phi_{\eta'\sigma j}\left(\mathbf{k}\right)\delta\left(\epsilon-\epsilon_{j}\left(\mathbf{k}\right)\right),\label{eq:4p_dos_mat}\end{equation}
 where $\phi_{\eta j}\left(\mathbf{k}\right)$ is the amplitude of
the Cu $p_{\eta}$ component in the band state specified by the band
index $j$ and momentum $\mathbf{k}$ with eigenenergy $\epsilon_{j}\left(\mathbf{k}\right)$.
The off-diagonal components ($\eta\ne\eta'$) are negligible because
of symmetry. This expression comes from the upper triangle in Fig.\ref{fig:diagram1}.

The factor $Y_{d}^{+-}\left(q\right)$ in Eq.~(\ref{eq.general})
is the density-density correlation function of the Keldysh type, which
is defined by \begin{equation}
Y_{d}^{+-}({\bf q},\omega)=\sum_{\lambda m\sigma}\sum_{\lambda^{\prime}m^{\prime}\sigma^{\prime}}Y_{\lambda m\sigma,\lambda^{\prime}m^{\prime}\sigma^{\prime}}^{+-}\left(q\right),\end{equation}
 where \begin{equation}
Y_{\lambda'm'\sigma',\lambda m\sigma}^{+-}({\bf q},\omega)=\int_{-\infty}^{\infty}\langle(\rho_{{\bf q}\lambda'm'\sigma'})^{\dagger}(\tau)\rho_{{\bf q}\lambda m\sigma}(0)\rangle{\rm e}^{i\omega\tau}{\rm d}\tau,\label{eq.y+-}\end{equation}
 with \begin{equation}
\rho_{{\bf q}\lambda m\sigma}=\sqrt{\frac{2}{N}}\sum_{{\bf k}}d_{{\bf k+q}\lambda m\sigma}^{\dagger}d_{{\bf k}\lambda m\sigma}.\end{equation}
 Here ${\bf k}$ runs over the magnetic first BZ. The index $\lambda m\sigma$
specifies a tight-binding orbital at site $\lambda$ with orbital
$m$ and spin $\sigma$. We assign $\lambda=1,2$ to two Cu sites
in the unit cell.

Using the solution within the HFA, we obtain \begin{eqnarray}
Y_{\lambda m\sigma,\lambda^{\prime}m^{\prime}\sigma^{\prime}}^{+-(0)}\left(q\right) & = & \Pi_{\lambda m\sigma\lambda m\sigma,\lambda^{\prime}m^{\prime}\sigma^{\prime}\lambda^{\prime}m^{\prime}\sigma^{\prime}}^{+-(0)}\left(q\right),\label{eq:Y-HFA}\end{eqnarray}
 with \begin{eqnarray}
 &  & \Pi_{\lambda_{1}m_{1}\sigma_{1}\lambda_{2}m_{2}\sigma_{2},\lambda_{1}^{\prime}m_{1}^{\prime}\sigma_{1}^{\prime}\lambda_{2}^{\prime}m_{2}^{\prime}\sigma_{2}^{\prime}}^{+-(0)}\left(q\right)\nonumber \\
 & = & \frac{2\pi}{N}\sum_{{\bf k}}\sum_{j,j'}\delta\left(\omega-E_{j'}\left(\mathbf{k}+\mathbf{q}\right)+E_{j}\left(\mathbf{k}\right)\right)\left[1-n_{j'}\left(\mathbf{k}+\mathbf{q}\right)\right]n_{j}\left(\mathbf{k}\right)\nonumber \\
 & \times & \varphi_{\lambda_{1}m_{1}\sigma_{1},j'}\left(\mathbf{k}+\mathbf{q}\right)\varphi_{\lambda_{1}^{\prime}m_{1}^{\prime}\sigma_{1}^{\prime},j'}^{*}\left(\mathbf{k}+\mathbf{q}\right)\varphi_{\lambda_{2}^{\prime}m_{2}^{\prime}\sigma_{2}^{\prime},j}\left(\mathbf{k}\right)\varphi_{\lambda_{2}m_{2}\sigma_{2},j}^{*}\left(\mathbf{k}\right).\label{eq.keldysh1}\end{eqnarray}
 where $E_{j}\left(\mathbf{k}\right)$ and $n_{j}\left(\mathbf{k}\right)$
are the eigenenergy and the occupation number of the eigenstate specified
by $j\mathbf{k}$, respectively. The $\varphi_{\lambda m\sigma,j}\left(\mathbf{k}\right)$
represents the amplitude of the $3d$ orbital and spin $m\sigma$
at the Cu site $\lambda$ in the energy eigenstate specified by $j\mathbf{k}$.
This expression describes the excitation from occupied states to unoccupied
states with Cu $3d$ amplitudes.

We treat the correlation effect on the electron-hole pair by the RPA.
It is included into a vertex function: \begin{eqnarray}
 &  & Y_{\xi,\xi^{\prime}}^{+-}\left(q\right)=\sum_{\xi_{1}\xi_{2}}\sum_{\xi_{1}^{\prime}\xi_{2}^{\prime}}\Lambda_{\xi_{1}\xi_{2},\xi}^{*}(q)\Pi_{\xi_{1}\xi_{2},\xi_{1}^{\prime}\xi_{2}^{\prime}}^{+-(0)}\left(q\right)\Lambda_{\xi_{1}^{\prime}\xi_{2}^{\prime},\xi^{\prime}}\left(q\right).\label{eq:Y-RPA}\end{eqnarray}
 with the vertex \begin{equation}
\Lambda_{\xi_{1}\xi_{2},\xi}(q)=\left[\hat{I}-\hat{\Gamma}\hat{F}^{--}\left(q\right)\right]_{\xi_{1}\xi_{2},\xi\xi}^{-1}.\label{eq.vertex}\end{equation}
 To simplify the notation, we abbreviate the indices $\lambda m\sigma$
as $\xi$. In Eq.~(\ref{eq.vertex}), $\hat{I}$ represents a unit
matrix, and $\hat{\Gamma}$ is the bare four-point vertex given by
\begin{equation}
\left[\hat{\Gamma}\right]_{\xi_{1}\xi_{2},\xi_{3}\xi_{4}}=\Gamma^{(0)}\left(m_{1}\sigma_{1}m_{4}\sigma_{4};m_{2}\sigma_{2}m_{3}\sigma_{3}\right)\delta_{\lambda_{1}\lambda_{2}}\delta_{\lambda_{3}\lambda_{4}}\delta_{\lambda_{1}\lambda_{3}},\end{equation}
 with $\xi_{1}=\lambda_{1}m_{1}\sigma_{1}$, $\xi_{2}=\lambda_{2}m_{2}\sigma_{2}$,
$\xi_{3}=\lambda_{3}m_{3}\sigma_{3}$, $\xi_{4}=\lambda_{4}m_{4}\sigma_{4}$
($\lambda_{i}=1,2$). Note that $\hat{\Gamma}$ is non-zero only for
$\xi_{1}$, $\xi_{2}$, $\xi_{3}$, $\xi_{4}$ belonging to the \emph{same}
Cu site. The two-particle propagator $\hat{F}^{--}(q)$ is given by
\begin{eqnarray}
\left[\hat{F}^{--}\left(q\right)\right]_{\xi_{1}\xi_{2},\xi_{3}\xi_{4}} & = & \frac{1}{N}\sum_{{\bf k}}\varphi_{\lambda_{4}m_{4}\sigma_{4},j}\left(\mathbf{k}\right)\varphi_{\lambda_{2}m_{2}\sigma_{2},j}^{*}\left(\mathbf{k}\right)\varphi_{\lambda_{1}m_{1}\sigma_{1},j'}\left(\mathbf{k}+\mathbf{q}\right)\varphi_{\lambda_{3}m_{3}\sigma_{3},j'}^{*}\left(\mathbf{k}+\mathbf{q}\right)\nonumber \\
 & \times & \left[\frac{n_{j}\left(\mathbf{k}\right)\left[1-n_{j'}\left(\mathbf{k}+\mathbf{q}\right)\right]}{\omega-E_{j'}\left(\mathbf{k}+\mathbf{q}\right)+E_{j}\left(\mathbf{k}\right)+i\delta}-\frac{n_{j'}\left(\mathbf{k}+\mathbf{q}\right)\left[1-n_{j}\left(\mathbf{k}\right)\right]}{\omega-E_{j'}\left(\mathbf{k}+\mathbf{q}\right)+E_{j}\left(\mathbf{k}\right)-i\delta}\right].\label{eq.green_pair}\end{eqnarray}
 For the details of the derivation, see refs.\citen{Igarashi2006}
and \citen{Takahashi2007NiO}. The RPA correction has been
found to have an important role in the momentum dependence of the
spectra.

\section{Calculated Results\label{sec:Calculated-Results}}

\begin{figure}[t]
\begin{center}
 \includegraphics[clip,scale=0.7]{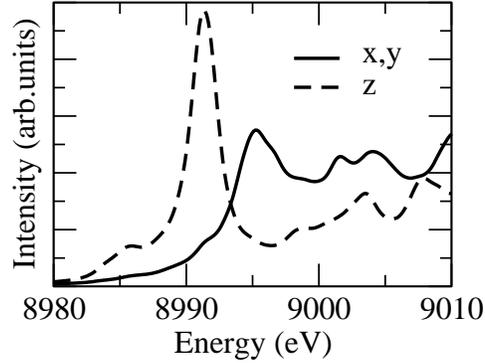}
\end{center}

\caption{\label{fig:dos-4p}The LDA $4p$ DOS convoluted by the Lorentzian
function with FWHM $2\Gamma_{1s}=1.6\,\mathrm{eV}$. In connection
with the absorption coefficient, the origin of energy is shifted so
that the prominent peak for the $p_{z}$ DOS locates at $8991\,\mathrm{eV}$.
A thin line represents the absorption coefficient with taking account
of the dipole matrix element evaluated by the band calculation.}
\end{figure}

In order to calculate the incident-photon-dependent factor $|L_{B}^{\eta}(\omega_{i};\omega)|^{2}$,
we need the $4p$ DOS, $\rho_{4p}^{\eta\eta}(\epsilon)$. We evaluate
the $4p$ DOS in a nonmagnetic (NM) phase on the basis of the muffin-tin
KKR band structure calculation method within the LDA. It is known
that the LDA fails to predict an insulating phase in the NM phase.
However, the $3d$ states forming a metallic phase hybridize little
with the $4p$ bands, giving rise to a minor effect on the $4p$ DOS
distributing around 5-30 eV above the Fermi level. Figure \ref{fig:dos-4p}
shows the $4p$ DOS convoluted with a Lorentzian function of FWHM
$2\Gamma_{1s}=1.6\,\mathrm{eV}$. Reflecting the layered structure,
$\rho_{4p}^{zz}(\epsilon)$ is quite different from $\rho_{4p}^{xx}(\epsilon)$.
Note that the $4p$ DOS is nearly proportional to the Cu $K$-edge
absorption spectra under the condition that the dipole matrix element
is constant and the interaction is neglected between the core hole
and the $4p$ electron. In connection with the absorption coefficient,
we set the energy difference between the Cu $1s$ level and the prominent
peak in the $4p_{z}$ DOS to be $8991\,\mathrm{eV}$.

\begin{figure}[t]
\begin{center}
 \includegraphics[clip,scale=0.8]{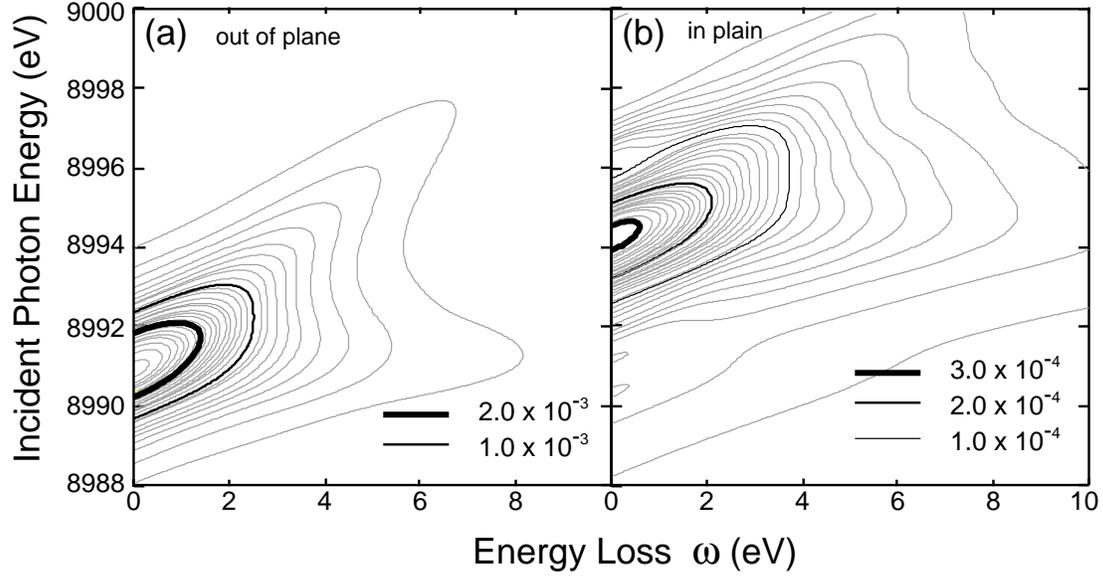}
\end{center}

\caption{\label{fig:LBC} Contour plot of $\left|L_{B}^{\eta}\left(\omega_{i},\omega\right)\right|^{2}$
as a function of energy loss $\omega$ and incident photon energy $\omega_i$ in the out-of-plane polarization
($\eta=z$) (a) and in the in-plane polarization ($\eta=x $) (b) with $\Gamma_{1s}=1$ eV.  Units of the intensity
are arbitrary.}
\end{figure}

Figure \ref{fig:LBC} shows the contour plot of $\left|L_{B}^{\eta}(\omega_{i};\omega)\right|^{2}$
calculated by substituting the $4p$ DOS into Eq.~(\ref{eq.born}).
The enhancement
starts earlier with increasing incident-photon energies in the out-of-plane
polarization ($\eta=z$), reflecting the difference between $\rho_{4p}^{zz}(\epsilon)$
and $\rho_{4p}^{xx}(\epsilon)$. Except for these points, the general
tendency of the enhancement factor looks similar in both polarizations;
a large peak appears first in the small energy loss region and moves
toward the higher energy loss region with increasing incident photon
energy. The overall intensities decrease with increasing incident
photon energy.

\begin{figure}[t]
\begin{center}
 \includegraphics[clip,scale=0.5]{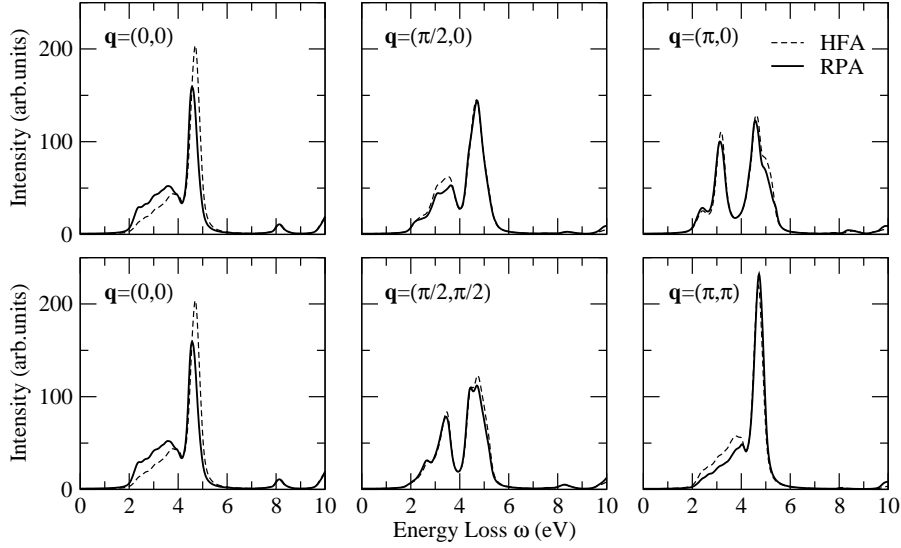}
\end{center}

\caption{\label{fig:Yd} The correlation function $Y_{d}^{+-}({\bf q},\omega)$
as a function of energy loss $\omega$. Top panels are for ${\bf q}$
along $(0,0)-(\pi,0)$, while bottom panels are for ${\bf q}$ along
$(0,0)-(\pi,\pi)$.}
\end{figure}

Another factor $Y_{d}^{+-}(q)$ mainly determines the RIXS spectra
as a function of energy loss. We calculate this factor from Eqs.~(\ref{eq:Y-HFA})
and (\ref{eq:Y-RPA}) with replacing the $\delta$ function by a Lorentzian
function ($\mathrm{FWHM}=0.2\,\mathrm{eV}$) in order to simulate
the instrumental resolution. Figure \ref{fig:Yd} shows the calculated
results for ${\bf q}$ along symmetry lines. We show the results with
$F^{0}=11.3\,\mathrm{eV}$ (corresponding to $U=10.7\,\mathrm{eV}$
in the $d$-$p$ model analysis \cite{Nomura2005}). We have checked
that another choice $F^{0}=7.3$ eV leads to essentially the same
result, except for the appearance of a weak intensity around $\omega=6\,\mathrm{eV}$.
We obtain continuous spectra ranging from $\omega=2$ eV to $6$ eV.
Intensities around $\omega=2$, $3.2$, and $4.5\,\mathrm{eV}$ are
caused by charge excitations of B$\rightarrow$A, C$\rightarrow$A,
and D$\rightarrow$A transitions, respectively, within the $x^{2}-y^{2}$
symmetry in the minority spin states (see Figs. \ref{fig:dos} and
\ref{fig:disp}).

We have a prominent peak around $4.5$ eV, which stays at the same
position with changing momentum. We also find that the spectral shape
in the low energy region changes with changing momentum; a broad hump
exists around $\omega=2-4$ eV at ${\bf q}=(0,0)$, which is enhanced
by the RPA correction. This hump grows up to become a peak around
$3.2$ eV with changing ${\bf q}$ along symmetry lines $(0,0)-(\pi,0)$
and $(0,0)-(\pi/2,\pi/2)$. Meanwhile, the $3.2$ eV peak is suppressed
with changing ${\bf q}$ along a symmetry line $(\pi/2,\pi/2)-(\pi,\pi)$,
vanishing at ${\bf q}=(\pi,\pi)$. The RPA correction help suppress
further the intensity in a region of $\omega=2-4$ eV at ${\bf q}=(\pi,\pi)$,
in contrast to the enhancement of the intensity at ${\bf q}=(0,0)$.

\begin{figure}[t]
\begin{center}
\includegraphics[clip,scale=1.00]{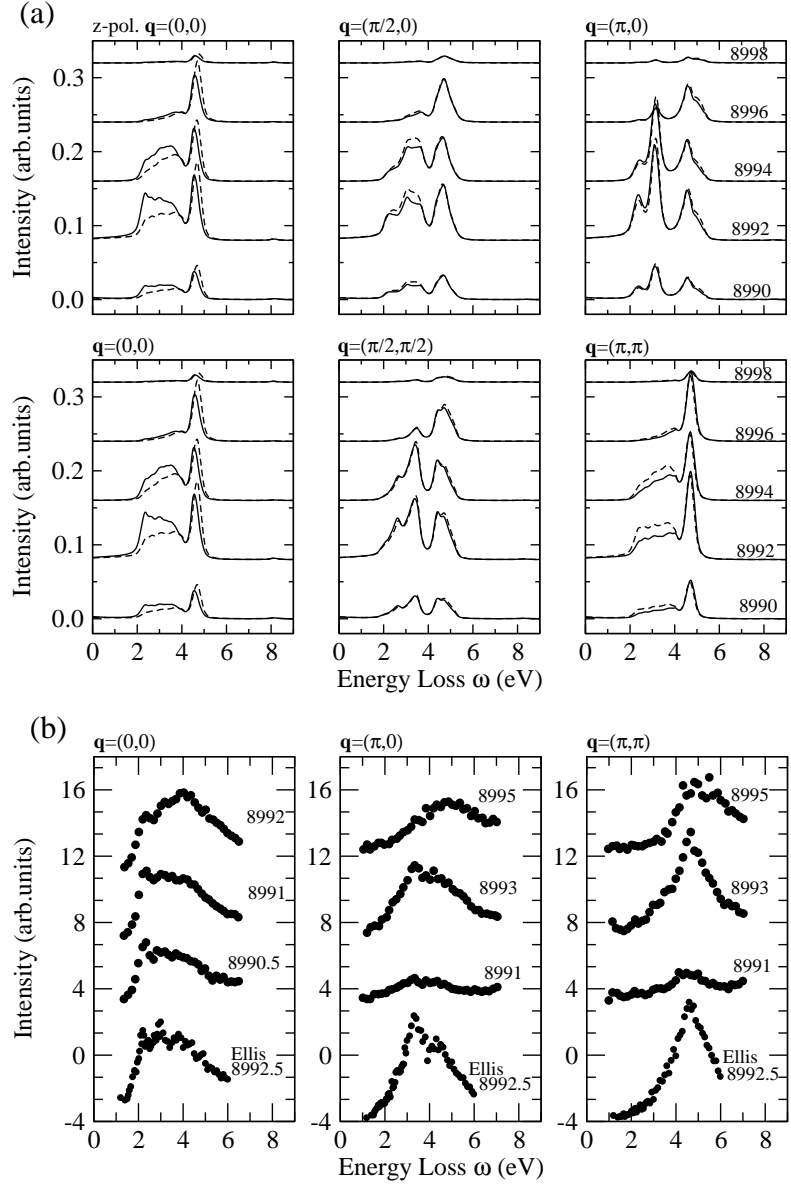}
\end{center}

\caption{\label{fig:RIXS-z} (a) RIXS spectra as a function
of energy loss in the out-of-plane ($z$) polarization. Top panels
are for ${\bf q}$ along a symmetric line $(0,0)-(\pi,0)$, while
bottom panels are for ${\bf q}$ along $(0,0)-(\pi,\pi)$. The solid and dashed
curves are the spectra calculated within the RP and HF approximations, respectively. (b) RIXS spectra
reproduced from refs. \citen{Lu2006,Ellis2007}. The curves denoted Ellis are taken from the ref. \citen{Ellis2007}. The numbers in panels
indicate the incident photon energy $\omega_{i}$.}
\end{figure}

The RIXS spectra are given by the product of $\left|L_{B}^{\eta}(\omega_{i},\omega)\right|^{2}$
and $Y_{d}^{+-}(q)$. Figure \ref{fig:RIXS-z} shows the calculated
spectra as a function of energy loss $\omega$ in the out-of-plane
polarization, in comparison with the experiments.\cite{Lu2006,Ellis2007} At ${\bf q}=(0,0)$,
a hump-like intensity, which is found around $\omega=2-4$ eV in $Y_{d}^{+-}({\bf q}=0,\omega)$,
is enhanced by $\left|L_{B}^{z}(\omega_{i},\omega)\right|^{2}$ with
$\omega_{i}=8990-8994$ eV. Since the enhancement is larger with lower
$\omega$, intensities around $\omega=2$ eV become a peak-like shape
in the RIXS spectra. Intensities around $\omega=3-4$ eV in $Y_{d}^{+-}({\bf q}=0,\omega)$
are moderately enhanced by $\left|L_{B}^{z}(\omega_{i},\omega)\right|^{2}$,
forming a very broad peak in the RIXS spectra. This broad spectral
feature, which is not produced by the $d$-$p$ model analysis,\cite{Nomura2005}
is consistent with the experiments.\cite{Lu2006,Kim2007,Ellis2007} The large $4.5$
eV peak in $Y_{d}^{+-}({\bf q}=0,\omega)$ loses its weight in comparison
with the intensities around $\omega=2-4$ eV because of small enhancement
by $\left|L_{B}^{z}(\omega_{i};\omega)\right|^{2}$. With increasing
$\omega_{i}$($>8994$ eV), the RIXS intensities around $\omega=2-4$
eV decrease, since the peak in $\left|L_{B}^{z}(\omega_{i};\omega)\right|^{2}$
as a function of energy loss $\omega$ moves to a region of higher $\omega$.
The intensity of the $4.5$ eV peak first changes little and then
decreases with further increasing $\omega_{i}$, since the $\left|L_{B}^{z}(\omega_{i},\omega)\right|^{2}$
becomes smaller.
With changing ${\bf q}$
along $(0,0)-(\pi,0)$, and $(0,0)-(\pi/2,\pi/2)$, a peak develops
around $\omega=3.2$ eV in accordance with the development of the peak
in $Y_{d}^{+-}({\bf q},\omega)$ for $\omega_{i}=8992-8994$ eV. The
intensity of the peak around $\omega=3.2$ eV becomes comparable to the $4.5$ eV peak at ${\bf q}=(\pi,0)$
and $(\pi/2,\pi/2)$, while intensities in the region of $\omega<3$ eV
become smaller in comparison with the corresponding spectra at ${\bf q}=(0,0)$.
This behavior corresponds well to the experimental line shape at ${\bf q}=(\pi,0)$,
which looks like two peaks around $\omega=3.2$ and $4.5$ eV.
With changing ${\bf q}$ along $(\pi/2,\pi/2)-(\pi,\pi)$, intensities
around $\omega=2-4$ eV are suppressed by the RPA correction in accordance
with the change of $Y_{d}^{+-}({\bf q},\omega)$, and only one peak
is overwhelmingly left at $4.5$ eV for a wide range of $\omega_{i}=8990-8996$
eV. This explains the experimental spectra at ${\bf q}=(\pi,\pi)$,
which looks like a single peak around $\omega=4.5$ eV. Although the
spectral shape is more consistent with the experiments\cite{Lu2006,Kim2007,Ellis2007} than that calculated
with the $d$-$p$ model on the whole, several discrepancies between the present calculation and the experiments
still remain; the intensity around $\omega=2$ eV is remarkably
enhanced at $\mathbf{q}=(0,0)$ with $\omega_{i}=8990.5$ eV, the $4.5$
eV peak is hardly discernible at $\mathbf{q}=(0,0)$, the intensity
at $\omega=4$ eV is enough large forming a peak-like structure
at $\mathbf{q}=(0,0)$ with $\omega_{i}=8992$ eV, and peak-like structures
are found in the higher energy loss region $\omega\gtrsim5$ eV in
the experiments.
We note that  peak structures are observed around $\omega=4.5$ eV at ${\bf q}\ne (0,0)$,
which seem to correspond well to the calculated $4.5$ eV peaks.
The calculated spectra at ${\bf q}=(0,0)$ specifically fail to reproduce the observed spectral structure in $\omega=4-4.5$ eV.
To remove these discrepancies, we may need to take
account of the electron correlations beyond the RPA and the effects
beyond the Born approximation to the core-hole potential.

\begin{figure}[t]
\begin{center}
\includegraphics[clip,scale=0.57]{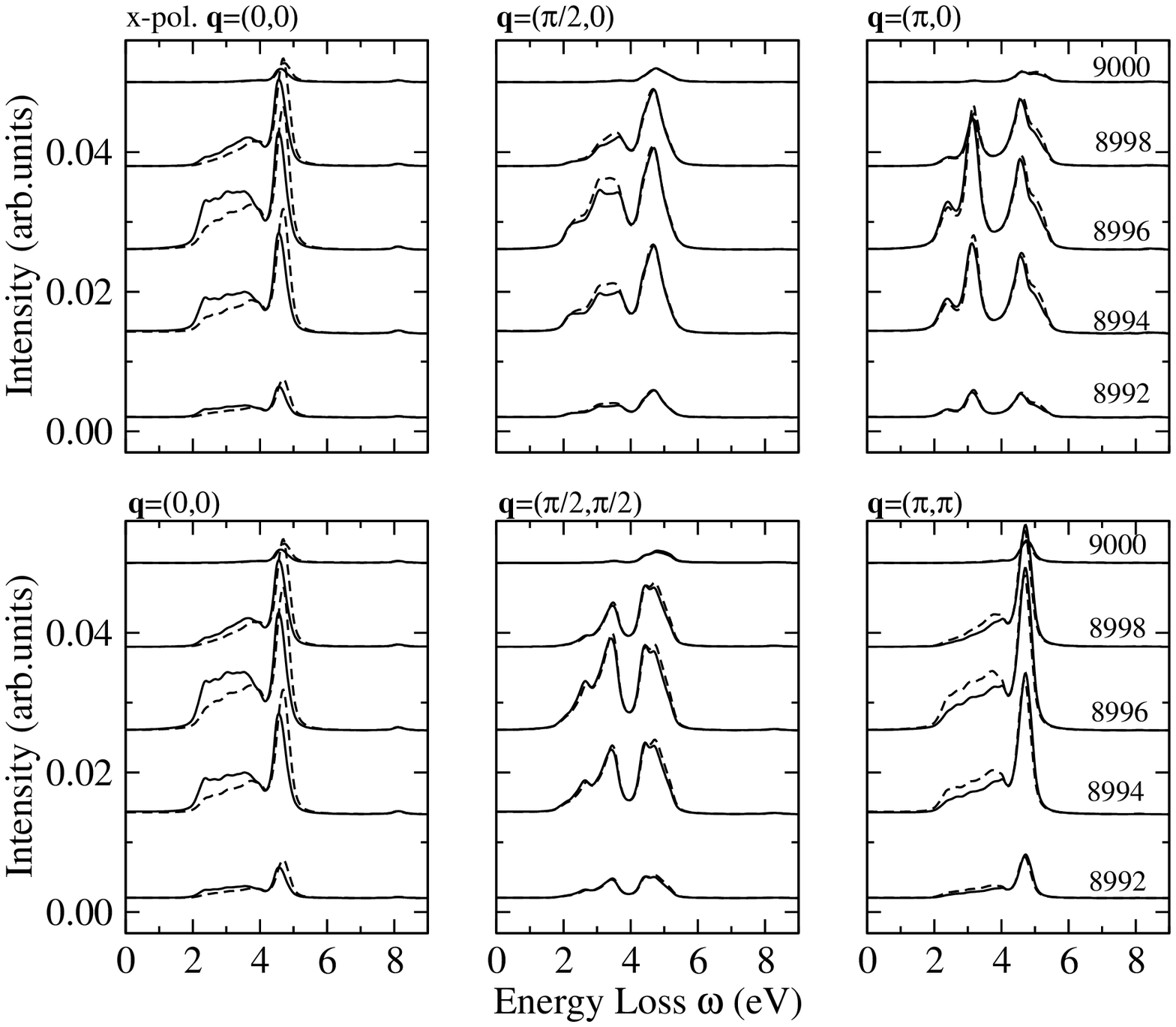}
\end{center}

\caption{\label{fig:RIXS-xy} RIXS spectra as a function
of energy loss for the in-plane ($xy$) polarization. Each panel is
for a typical ${\bf q}$ value. The solid and dashed curves are the spectra
calculated within the RP and HF approximations, respectively. The unit of intensity is the same as in
Fig.~\ref{fig:RIXS-z}.}

\end{figure}

Figure \ref{fig:RIXS-xy} shows the calculated RIXS spectra as a function
of energy loss $\omega$ in the in-plane polarization.
The scale of the calculated intensity is the
same as that in the out-of-plane polarization. The enhancement by
$\left|L_{B}^{x}(\omega_{i};\omega)\right|^{2}$ starts to be effective
at $\omega_{i}$ about $3$ eV higher with magnitude much smaller
compared to that by $\left|L_{B}^{z}(\omega_{i};\omega)\right|^{2}$.
Reflecting these differences, the RIXS spectra are enhanced for $\omega_{i}>8993$
eV with smaller intensities than those in the out-of-plane polarization.
Except for these points, the change in spectra with changing ${\bf q}$
and the incident-photon energy are similar to those in the out-of-plane
polarization. This is consistent with the recent experiment by Kim et al.\cite{Kim2007}

In comparison with our previous studies based on a $d$-$p$ model
with fixing $\omega_{i}$ at $8991$ eV in the out-of-plane polarization,
\cite{Igarashi2006} the spectral change obtained in the present paper
with changing ${\bf q}$ is consistent with our previous study. Difference
is that the present study using a multiorbital tight-binding model
leads to a distribution of intensities around $2-4$ eV.
This corresponds well to the experimental spectral shape.
The present study also improves peak position
from our previous value $\omega=5.5$ eV\cite{Nomura2005,Igarashi2006}
to $4.5$ eV, although the spectral shape still needs to be improved.

\section{Concluding Remarks\label{sec:Concluding-Remarks}}

We have analyzed the incident-photon-energy and polarization dependences
on the RIXS spectra in La$_{2}$CuO$_{4}$ on the basis of the formula
developed by Nomura and Igarashi. This formula expresses the RIXS
spectra by a product of the density-density correlation function and
the incident-photon dependent factor. To explain the fine structures
found in the RIXS experiment, we have extended the $d$-$p$ model
used in our previous analyses to a multiorbital tight-binding model,
which includes all the Cu $3d$ and O $2p$ orbitals as well as the
full Coulomb interaction between $3d$ orbitals. We have calculated
the density-density correlation function using the HFA and RPA, and
the incident-photon-dependent factor using the $4p$ DOS from the
ab initio band structure calculation. Fine structures are hardly reproduced
by the $d$-$p$ model. In the present detailed analysis, they are
found in energy loss $2\sim5$ eV, which vary with varying momentum
and incident-photon energy, in semi-quantitative agreement with the
experiment. Note that the peak shifts as a function of momentum should
not be interpreted as a dispersion relation of a kind of exciton,
because the peaks with broad widths constitute an energy continuum
generated by a band-to-band transition.
The experimental structures are not clear enough for more detailed comparison with
the calculated spectra, we hope the correspondence will be
clarified by improving the instrumental resolution.

The electron correlation works to modify the single-particle energy
bands given by the HFA.\cite{Imada1998} One prominent effect is a
creation of \char`\"{}satellite\char`\"{} peak around $6\sim11$ eV
below the top of the valence band. According to the three-body scattering
theory by the present authors, \cite{Taka1996} the creation of satellite
works to push the $3d$ states toward upper energy region in the valence
band. Such modifications are not strong on the minority spin $x^{2}-y^{2}$
states in the shallow energy region from the top of the valence band.
Another effect is the reduction of the energy gap from the HFA value.
We have taken account of this effect by adjusting the value of $\Delta$
to give the experimental energy gap within the HFA. Since the RIXS
spectra for $\omega<6$ eV come from the $3d$ states in the shallow
energy region of the valence band, these points may explain why the
HFA works rather well.
As regards the Born approximation to the core-hole potential, our calculation
explains semi-quantitatively the incident-photon energy dependence,
indicating that the Born approximation works rather well.
Although the present analysis is in semi-quantitative
agreement with the experiment, several drawbacks remain as pointed
out in the preceding section. Taking account of electron correlations
and going beyond the Born approximation to the core-hole potential
are desired for quantitative agreement with the experiment.\cite{Lu2006,Kim2007,Ellis2007}

Although experimental data have been accumulated for doped cuprates,
\cite{Ishii2005-1,Ishii2005-2,Lu2005} theoretical analyses are limited
on a one-band Hubbard model within the exact diagonalization method,\cite{Tsutsui2003}
and on a three-band Hubbard model analysis within the HFA on the basis
of the present formalism.\cite{Markiewicz2006} An analysis with a
detailed model like the present paper may be necessary to clarify
the momentum and incident-photon dependences of the spectra. Since
electron correlations are expected to be more important in doped cuprates,
such studies seem rather hard and are left in future.

\section*{Acknowledgment}
We thank J. Mizuki for valuable discussions. This work was partially
supported by a Grant-in-Aid for Scientific Research from the Ministry
of Education, Culture, Sports, Science, and Technology, Japan. 


\begin{thebibliography}{99}

\bibitem{Hill1998}
J. Hill, C.-C. Kao, W. Caliebe, M. Matsubara, A. Kotani, J. Peng, and R.
  Greene:  Phys. Rev. Lett. \textbf{80} (1998) 4976.
\bibitem{Abbamonte1999}
P. Abbamonte, C.~A. Burns, E.~D. Isaacs, P.~M. Platzman, L.~L. Miller, S.~W.
  Cheong, and M.~V. Klein:  Phys.\ Rev.\ Lett. \textbf{83} (1999) 860.
\bibitem{Hasan2000}
M. Hasan, E. Isaacs, Z.-X. Shen, L.~L. Miller, L. Tsutsui, T. Tohyama, and S.
  Maekawa:  Science \textbf{288} (2000) 1811.
\bibitem{Hasan2002}
M.~Z. Hasan, P.~A. Montano, E.~D. Isaacs, Z.-X. Shen, H. Eisaki, S.~K. Sinha,
  Z. Islam, N. Motoyama, and S. Uchida:  Phys. Rev. Lett. \textbf{88} (2002)
  177403.
\bibitem{Kim2002}
Y.~J. Kim, J.~P. Hill, C.~A. Burns, S. Wakimoto, R.~J. Birgeneau, D. Casa, T.
  Gog, and C.~T. Venkataraman:  Phys. Rev. Lett. \textbf{89} (2002) 177003.
\bibitem{Inami2003}
T. Inami, T. Fukuda, J. Mizuki, S. Ishihara, H. Kondo, H. Nakao, T. Matsumura,
  K. Hirota, Y. Murakami, S. Maekawa, and Y. Endoh:  Phys. Rev. B \textbf{67}
  (2003) 045108.
\bibitem{Kim2004PRB}
Y.-J. Kim, J.~P. Hill, S. Komiya, Y. Ando, D. Casa, T. Gog, and C.~T.
  Venkataraman:  Phys. Rev. B \textbf{70} (2004) 094524.
\bibitem{Kim2004PRL}
Y.-J. Kim, J.~P. Hill, H. Benthien, F.~H.~L. Essler, E. Jeckelmann, H.~S. Choi,
  T.~W. Noh, N. Motoyama, K.~M. Kojima, S. Uchida, D. Casa, and T. Gog:  Phys.
  Rev. Lett. \textbf{92} (2004) 137402.
\bibitem{Suga2005}
S. Suga, S. Imada, A. Higashiya, A. Shigemoto, S. Kasai, M. Sing, H. Fujiwara,
  A. Sekiyama, A. Yamasaki, C. Kim, T. Nomura, J. Igarashi, M. Yabashi, and T.
  Ishikawa:  Phys. Rev. B \textbf{72} (2005) 081101.
\bibitem{Ishii2005-1}
K. Ishii, K. Tsutsui, Y. Endoh, T. Tohyama, K. Kuzushita, T. Inami, K. Ohwada,
  S. Maekawa, T. Masui, S. Tajima, Y. Murakami, and J. Mizuki:  Phys. Rev.
  Lett. \textbf{94} (2005) 187002.
\bibitem{Ishii2005-2}
K. Ishii, K. Tsutsui, Y. Endoh, T. Tohyama, S. Maekawa, M. Hoesch, K.
  Kuzushita, M. Tsubota, T. Inami, J. Mizuki, Y. Murakami, and K. Yamada:
  Phys. Rev. Lett. \textbf{94} (2005) 207003.
\bibitem{Lu2005}
L. Lu, G. Chabot-Couture, X. Zhao, J. Hancock, N. Kaneko, O. Vajk, G. Yu, S.
  Grenier, Y.-J. Kim, D. Casa, and M. Greven:  Phys. Rev. Lett. \textbf{95} (2005) 217003.
\bibitem{Lu2006}
L. Lu, J.~N. Hancock, G. Chabot-Couture, K. Ishii, O.~P. Vajk, G. Yu, J.
  Mizuki, D. Casa, T. Gog, and M. Greven:  Phys. Rev. B \textbf{74} (2006) 224509.
\bibitem{Kim2007}
Y.-J. Kim, J. P. Hill, S. Wakimoto, R. J. Birgeneau, F. C. Chou, N. Motoyama, K. M. Kojima, S. Uchida, D. Casa, and T. Cog:
 Phys. Rev. B \textbf{76} (2007) 155116.
\bibitem{Ellis2007}
D.S. Ellis, J. P. Hill, S. Wakimoto, R. J. Birgeneau, D. Casa, T. Gog, Y.-J. Kim: Phys. Rev. B \textbf{77} (2008) 060501.
\bibitem{Collart2006}
E. Collart, A. Shukla, J.-P. Rueff, P. Leininger, H. Ishii, I. Jarrige, Y.~Q.
  Cai, S.-W. Cheong, and G. Dhalenne:  Phys. Rev. Lett. \textbf{96} (2006) 157004.
\bibitem{Tsutsui1999}
K. Tsutsui, T. Tohyama, and S. Maekawa:  Phys. Rev. Lett. \textbf{83} (1999) 3705.
\bibitem{Tsutsui2000}
K. Tsutsui, T. Tohyama, and S. Maekawa:  Phys. Rev. B \textbf{61} (2000) 7180.
\bibitem{Nomura2004}
T. Nomura and J.-i. Igarashi:  J. Phys. Soc. Jpn. \textbf{73} (2004) 1677.
\bibitem{Nomura2005}
T. Nomura and J.-i. Igarashi:  Phys. Rev. B \textbf{71} (2005) 035110.
\bibitem{Okada2006}
K. Okada and A. Kotani:  J. Phys. Soc. Jpn. \textbf{75} (2006) 044702.
\bibitem{Igarashi2006}
J.-i. Igarashi, T. Nomura, and M. Takahashi:  Phys. Rev. B \textbf{74} (2006) 245122.
\bibitem{Nozieres1974}
P. Nozi\`eres and E. Abrahams:  Phys. Rev. B \textbf{10} (1974) 3099.
\bibitem{Brink2006}
J. van~den Brink and M. van Veenendaal:  Euro. Phys. Lett. \textbf{73} (2006) 121.
\bibitem{Nozieres1969_1097}
P. Nozi{\`e}res and C.~T.~D. Dominicis:  Phys. Rev. \textbf{178} (1969) 1097.
\bibitem{Takahashi2007NiO}
M. Takahashi, J. Igarashi, and T. Nomura: Phys. Rev. \textbf{75} (2007) 235113.
\bibitem{H.Ishii2006} H. Ishii: private communication.
\bibitem{DeWeert1989}
M.~J. DeWeert, D.~A. Papaconstantopoulos, and W.~E. Pickett:  Phys. Rev. B \textbf{39} (1989) 4235.
\bibitem{Slater1954}
J.~C. Slater and G.~F. Koster:  Phys. Rev. \textbf{94} (1954) 1498.
\bibitem{Eskes1991}
H. Eskes and G.~A. Sawatzky:  Phys. Rev. B \textbf{43} (1991) 119.
\bibitem{Mizokawa1996}
T. Mizokawa and A. Fujimori:  Phys. Rev. B \textbf{53} (1996) 4201(R).
\bibitem{Takahashi1999}
M. Takahashi, J. Igarashi, and P. Fulde:  J. Phys. Soc. Jpn. \textbf{68} (1999) 2530.
\bibitem{Imada1998}
M. Imada, A. Fujimori, and Y. Tokura:  Rev. Mod. Phys. \textbf{70} (1998) 1039.
\bibitem{Taka1996}
M. Takahashi and J.-i. Igarashi:  Phys. Rev. B \textbf{54} (1996) 13566.
\bibitem{Tsutsui2003}
K. Tsutsui, T. Tohyama, and S. Maekawa:  Phys. Rev. Lett. \textbf{91} (2003) 117001.
\bibitem{Markiewicz2006}
R.~S. Markiewicz and A. Bansil:  Phys. Rev. Lett. \textbf{96} (2006) 107005.
\end{thebibliography}


%
\newpage

\end{document}